\documentstyle[iopconf1]{article}
\newcommand{\rf}[1]{(\ref{#1})}
\newcommand{\be}[1]{\begin{equation} \label{#1}}
\newcommand{\ee}{\end{equation}}
\newcommand{\ba}[1]{\begin{eqnarray} \label{#1}}
\newcommand{\ea}{\end{eqnarray}}
\newcommand{\ban}{\begin{eqnarray*}}
\newcommand{\ean}{\end{eqnarray*}}
\newcommand{\nn}{\nonumber}
\def\CN{{\cal N}}
\begin{document}
\title{Next-to-Minimal Supersymmetric Standard Model and 
	direct Dark Matter detection}

\author{V A~Bednyakov\dag\footnote{E-mail:
	bedny@nusun.jinr.ru} 
	and H V~Klapdor-Kleingrothaus\ddag\footnote{E-mail:
	Klapdor@mickey.mpi-hd.mpg.de}}

\affil{\dag\ 
        Laboratory of Nuclear Problems,
	Joint Institute for Nuclear Research,
        Moscow region, 141980 Dubna, Russia}

\affil{\ddag\ Max-Planck-Institut f\"{u}r Kernphysik, 
        Postfach 103980, D-69029, Heidelberg, Germany}

\beginabstract
        Prospects for direct dark matter detection are considered in the 
	Next-to-Minimal Supersymmetric Standard Model (NMSSM).
	The effective neutralino-quark Lagrangian is obtained and 
	event rates are calculated for $^{73}$Ge.
	Accelerator and cosmological constraints on the NMSSM
	are included in our numerical analysis,
	which allows us to conclude that 
	the lightest neutralino from NMSSM
	could be detected in future dark matter experiments.
\endabstract

\section{Introduction}
        The lightest supersymmetric particle (LSP), the neutralino,
        is considered now as most promising candidate, 
	which may contribute the main fraction of so-called cold dark matter.

        In this paper we investigate direct detection of the relic LSP in 
        the Next-To-Minimal Supersymmetric Standard Model (NMSSM) 
\cite{Drees, EGHRZ}.
        The NMSSM is mainly motivated by its potential to eliminate the
	so-called $\mu$ problem 
\cite{KimNillis} of the Minimal Supersymmetric Standard Model (MSSM) 
\cite{HaberKane}, where the origin of the 
	$\mu$ parameter in the superpotential
$ W_{\mbox{\scriptsize MSSM}}=\mu  H_1 H_2 $ is not understood.
	For phenomenological reasons it has to be of the order
	of the electroweak scale, while the "natural" mass scale would be
	of the order of the GUT or Planck scale.
	This problem is evaded in the NMSSM where the
	$\mu$ term in the superpotential is dynamically generated through
	$\mu = \lambda x$ with a dimensionless coupling $\lambda$ and
	the vacuum expectation value $x$ of the Higgs singlet.
        Another essential feature of the NMSSM is the fact that the 
	mass bounds for the Higgs bosons and neutralinos are weakened.
	While in the MSSM experimental data imply a lower mass bound
        of about 20 GeV for the LSP 
\cite{ALEPH}, 
	very light or massless neutralinos and Higgs bosons
	are not excluded in the NMSSM 
\cite{FrankeFraasBartl,FrankeFraas1}.
	The above arguments make an intensive study of the
	NMSSM phenomenology very desirable.
	Previously the Higgs and neutralino sectors
	of the NMSSM were carefully studied in
\cite{FrankeFraasBartl}%
--\cite{Pandita}.
	The calculation of the LSP relic abundance 
	in the NMSSM was performed for the first time in
\cite{ApelSW} and recently in  
\cite{AStephan}.
\smallskip

	The outline of this paper is as follows.
	In 
Sec.~\ref{sec:Lag} we describe the Lagrangian of the NMSSM.
Section~\ref{sec:dark} collects formulas relevant for 
	calculation of the event rate for direct dark matter 
	detection in the framework of the NMSSM.
	In 
Sec.~\ref{sec:constraints} we discuss the constraints on 
	the NMSSM parameter space which are used in our analysis.
	In 
Sec.~\ref{sec:numerical} we describe our numerical 
	procedure and discuss the results obtained.
Sec.~\ref{sec:conclusion} contains our conclusion.

\section{The Lagrangian of the NMSSM} \label{sec:Lag}
        The NMSSM superpotential is 
\cite{FrankeFraas} 
	($\varepsilon_{12}=-\varepsilon_{21}=1$):
\begin{eqnarray}
W & = & \lambda \varepsilon_{ij} H_1^i H_2^j N -\frac{1}{3} k N^3
\nonumber \\ & &
+h_u \varepsilon_{ij} \tilde{Q}^i \tilde{U} H_2^j
-h_d \varepsilon_{ij} \tilde{Q}^i \tilde{D} H_1^j
-h_e \varepsilon_{ij} \tilde{L}^i \tilde{R} H_1^j,
\end{eqnarray}
        where $H_1 = (H_1^0,H^-)$ and
	      $H_2=(H^+,H_2^0)$ are the SU(2) Higgs
	doublets with hypercharge $-1/2$ and $1/2$ and
	$N$ is the Higgs singlet with hypercharge 0.
	The notation of the fermion doublets and singlets
	is conventional, generation indices are omitted. 
	The electroweak gauge-symmetry 
	SU(2)$_{\rm I}\times$U(1)$_{\rm Y}$
	is spontaneously broken to the electromagnetic
	gauge-symmetry U(1)$_{\rm em}$ by the Higgs VEVs
	$\langle H_i^0 \rangle = v_i$  with $i = 1,2$
	and $\langle N \rangle = x$,
	where $v = \sqrt{v_1^2 + v_2^2} = 174$~GeV,
	$\tan \beta = v_2/v_1$.

        The most general supersymmetry breaking potential
	has the from 
\cite{FrankeFraas}
\begin{eqnarray}
\label{vs}
V_{\mbox{soft}} & = &
m_1^2 |H_1|^2 + m_2^2 |H_2|^2+m_3^2 |N|^2
\nonumber \\ & &
+m_Q^2 |\tilde{Q}|^2 + m_U^2 |\tilde{U}|^2 + m_D^2 |
\tilde{D}|^2
+m_L^2 |\tilde{L}|^2 + m_E^2 |\tilde{R}|^2
\nonumber \\  & &
-(\lambda A_\lambda \varepsilon_{ij} H_1^i H_2^j N + \mbox{h.c.})
-(\frac{1}{3}kA_k N^3 + \mbox{h.c.})
\nonumber \\  & &
+(h_u A_U \varepsilon_{ij} \tilde{Q}^i \tilde{U} H_2^j
-h_d A_D \varepsilon_{ij} \tilde{Q}^i \tilde{D} H_1^j
-h_e A_E \varepsilon_{ij} \tilde{L}^i \tilde{R} H_1^j
+\mbox{h.c.})
\nonumber \\ & &
+\frac{1}{2}M \lambda^a \lambda^a
+\frac{1}{2}M' \lambda ' \lambda '.
\end{eqnarray}

	The minimization conditions for the scalar potential
$\partial V / \partial v_{1,2} =0$ and 
$\partial V / \partial x =0$
	eliminate three parameters of the Higgs sector
	which are normally chosen to be $m_1^2$, $m_2^2$ and $m_3^2$.

        Therefore at the weak scale we are left with
	the following free parameters of the NMSSM.
	There are
	the ratio of the doublet vacuum expectation values,
	$\tan\beta$, the singlet vacuum expectation value $x$,
	the couplings in the superpotential $\lambda$ and $k$,
	the parameters $A_\lambda$, $A_k$,
	as well as $A_U$, $A_D$, $A_E$ (for three generations)
	in the supersymmetry breaking potential, the gaugino mass
	parameters $M$ and $M'$, and the scalar mass parameters
	for the squarks $m_{Q,U,D}$ and sleptons $m_{L,E}$.

	Assuming CP conservation in the Higgs sector,
	the Higgs CP-even and CP-odd 
	$3\times 3$ matrices are diagonalized by the real
	orthogonal $3\times 3$ 
	matrices $U^S$ and $U^P$, respectively,
\begin{eqnarray*}
\mbox{Diag}(m_{S_1}^2,m_{S_2}^2,m_{S_3}^2) & = & {U^S}^T {\cal M}_S^2
U^S, \\
\mbox{Diag}(m_{P_1}^2,m_{P_2}^2,0) & = & {U^P}^T {\cal M}_P^2 U^P,
\end{eqnarray*}
	where $m_{S_1} < m_{S_2} < m_{S_3}$ and
	$m_{P_1}<m_{P_2}$ denote the masses of  
	the mass eigenstates of the neutral scalar 
	Higgs bosons $S_a$ $(a=1,2,3)$ 
	and neutral pseudoscalar Higgs bosons $P_\alpha$ $(\alpha = 1,2)$
\cite{FrankeFraas}.
	The Higgs sector of the NMSSM contains
	five physical neutral Higgs bosons, three Higgs scalars,
	two pseudoscalars, and two
	degenerate physical charged Higgs particles $C^\pm$.
	In our numerical analysis 
	we have included 1-loop radiative corrections
	to all Higgs mass matrices in accordance with 
\cite{ElliottKingWhite,KingWhite-rc}.

	The masses and mixings of the neutralinos in the basis 
$   \Psi^T = (-i \lambda_1,-i \lambda^3_2,
   \Psi^0_{H_1},\Psi^0_{H_2},\Psi_N)$
	are determined by the Lagrangian
\cite{FrankeFraas}
\begin{eqnarray*}
   {\cal L} = -\frac{1}{2} \Psi^T M \Psi + {\rm h.c.}.
\end{eqnarray*}
	The mass of the neutralinos is obtained by diagonalizing
	the 5$\times$5 matrix $M$ with the orthogonal matrix $N$:
\begin{eqnarray*}
   {\cal L} = -\frac{1}{2} \: m_i \:
   \overline{\tilde{\chi}^0_i} \: \tilde{\chi}^0_i,
\qquad
   \tilde{\chi}^0_i = \left(\begin{array}{cc}
   \chi^0_i  \\ \overline{\chi}^0_i
   \end{array}\right),
\end{eqnarray*}
	with
	$ \chi^0_i = {\cal N}_{ij} \Psi_j $ 
	and
   	${\rm diag} \left\{ m_1, m_2, ..., m_5\right\} 
	= N \: M \: N^{T}.$
	The neutralinos $\tilde{\chi}^0_i$ ($i$ = 1--5) are ordered
	with increasing mass $|m_i|$, thus $\chi\equiv
	\tilde{\chi}^0_1$ is the LSP neutralino.
	The matrix elements ${\cal N}_{i j}$ ($i,j$ = 1--5)
	describe the composition of the neutralino $\tilde{\chi}^0_i$ 
	in the basis $\Psi_j$.
	For example the bino fraction of the lightest neutralino is given
	by ${\cal N}^2_{1 1}$ and the singlino fraction of this neutralino
	by ${\cal N}^2_{1 5}$.
	The remaining particle content is identical with that of the MSSM.

\section{Neutralino-nucleus elastic scattering} \label{sec:dark}
        A dark matter event is elastic scattering of a DM neutralino from
   	a target nucleus producing a nuclear recoil which can be detected
	by a suitable detector.
        The corresponding event rate depends on the distribution of
        the DM neutralinos in the solar vicinity and
        the cross section of  neutralino-nucleus elastic scattering.

        An experimentally observable quantity is the differential 
	event rate per unit mass of the target material
$$
\frac{\d R}{\d E_r} = 
	        \Bigl[ N \frac{\rho_\chi}{m_\chi} \Bigr]
    \int^{v_{\rm max}}_{v_{\rm min}} \d v f(v) v 
	\frac{\d\sigma}{\d q^2} (v, E_r),
	\qquad q^2 = 2 M_A E_r.
$$
   	Here $f(v)$ is the velocity distribution of neutralinos
   	in the earth's frame which is usually assumed to be a
	Maxwellian distribution in the galactic frame.
$v_{\rm max} = v_{\rm esc} \approx$ 600 km/s and
$\rho_{\chi}$ = 0.3 GeV$\cdot$cm$^{-3}$  are the escape velocity
	and the mass density of the relic neutralinos in
   	the solar vicinity;
$v_{\rm min} = \left(M_A E_r/2 M_{\rm red}^2\right)^{1/2}$ with
$M_A$ and $M_{\rm red}$ being the mass of nucleus $A$ and the reduced
   	mass of the neutralino-nucleus system, respectively.

   	The differential event rate is the most appropriate quantity
	for comparing with the observed recoil spectrum and allows one
	to take properly into account spectral characteristics of a
	specific detector and to separate the background.
   	However, in many cases the total  event rate $R$\
   	integrated over the whole kinematical domain of the recoil energy
   	is sufficient.
   	It is widely employed in theoretical papers 
\cite{JungKamGriest}--\cite{Superlight},
\cite{GondoloBerg,BBEFMS},
	for estimating the
	prospects for DM detection,
   	ignoring experimental complications which may occur on the  way.
        In the present paper we use the total
	event rate $R$, because we
	are going  to perform a general
	analysis aimed at searching for domains with 
	large values of the event rate $R$.

        A general representation of the differential cross section
        of neutralino-nucleus scattering can be given in terms of
        three spin-dependent ${\cal  F}_{ij}(q^2)$ and
        one spin-independent ${\cal F}_{S}(q^2)$ form factors as follows
\cite{EngelVogel}
\ban 
\frac{\d\sigma}{\d q^2}
(v,q^2)&=&\frac{8 G_{\rm F}}{v^2} \Bigl(
   		a_0^2\cdot {\cal F}_{00}^2(q^2)
	     + a_0 a_1 \cdot {\cal F}_{10}^2(q^2) \nn \\
	     &+&
   	        a_1^2\cdot {\cal F}_{11}^2(q^2)
   	      + c_0^2\cdot A^2\ {\cal F}_{S}^2(q^2)
   	\Bigr).
\ean
   	The last term corresponding to the spin-independent scalar
	interaction gains coherent enhancement $A^2$
	($A$ is the atomic weight of the nucleus in the reaction).
   	The coefficients $a_{0,1}, c_0$ do not depend on nuclear structure
   	but relate to the the nucleon structure
	and the parameters ${\cal A}_q$, ${\cal C}_q$
	of the relevant low-energy effective neutralino-quark 
	Lagrangian 
\cite{JungKamGriest}--\cite{NathArnowitt}
\be{Lagr}
  L_{\rm eff} = \sum_{q}^{}\left( {\cal A}_{q}\cdot
      \bar\chi\gamma_\mu\gamma_5\chi\cdot
                \bar q\gamma^\mu\gamma_5 q +
    \frac{m_q}{M_{W}} \cdot{\cal C}_{q}\cdot\bar\chi\chi\cdot\bar q q\right),
\ee
        where terms with vector and pseudoscalar quark currents are
        omitted being negligible in the case of non-relativistic
        DM neutralinos with typical velocities $v \approx 10^{-3} c$.
	The coefficients in Lagrangian
\rf{Lagr} have the form:
\ban 
 {\cal A}_{q} =
	&-&\frac{g_{2}^{2}}{4M_{W}^{2}}
	   \left[ \frac{{\cal N}_{14}^2-{\cal N}_{13}^2}{2}T_3 \right . \\
        &-& \frac{M_{W}^2}{m^{2}_{\tilde{q}1} - (m_\chi + m_q)^2}
	   (\cos^{2}\theta_{q}\ \phi_{qL}^2
	   + \sin^{2}\theta_{q}\ \phi_{qR}^2)
\\      &-& \frac{M_{W}^2}{m^{2}_{\tilde{q}2} - (m_\chi + m_q)^2}
	     (\sin^{2}\theta_{q}\ \phi_{qL}^2
	     + \cos^{2}\theta_{q}\ \phi_{qR}^2) \nn \\
        &-& \frac{m_{q}^{2}}{4}P_{q}^{2}\left(\frac{1}{m^{2}_{\tilde{q}1}
		- (m_\chi + m_q)^2}
             + \frac{1}{m^{2}_{\tilde{q}2}
                - (m_\chi + m_q)^2}\right) 
\\      &-& \frac{m_{q}}{2}\  M_{W}\  P_{q}\  \sin2\theta_{q}\
            T_3 ({\cal N}_{12} - \tan\theta_W {\cal N}_{11}) 
\\     &\times& \left .
	\left( \frac{1}{m^{2}_{\tilde{q}1}- (m_\chi + m_q)^2}
    - \frac{1}{m^{2}_{\tilde{q}2} - (m_\chi + m_q)^2}\right)
	\right],
\ean
\ban
 {\cal C}_{q} =
	&-&  \frac{g_2^2}{4} 
	\left[
         - \sum^{}_{a=1,2,3} Q^{L\prime\prime}_{a11}
                 \frac{1}{m^2_a} 
           \Bigl[
                (\frac12 +T^{}_{3q}) \frac{U^S_{a2}}{\sin\!\beta}
               +(\frac12 -T^{}_{3q}) \frac{U^S_{a1}}{\cos\!\beta}
           \Bigr]
	\right. 
\\  &+& P_q \left(\frac{\cos^{2}\theta_{q}\ \phi_{qL} -
     \sin^{2}\theta_{q}\ \phi_{qR}}{m^{2}_{\tilde{q}1} - (m_\chi + m_q)^2}
         -\frac{\cos^{2}\theta_{q}\ \phi_{qR} -
     \sin^{2}\theta_{q}\ \phi_{qL}}{m^{2}_{\tilde{q}2} - (m_\chi +
	m_q)^2}\right) \\
	&+& \sin2\theta_{q}(\frac{m_q}{4 M_W} P_{q}^{2} -
                           \frac{M_W}{m_q} \phi_{qL}\ \phi_{qR})
\\   &\times& \left. 
	\left(\frac{1}{m^{2}_{\tilde{q}1} - (m_\chi + m_q)^2} -
\frac{1}{m^{2}_{\tilde{q}2} - (m_\chi + m_q)^2}\right)
	\right].
\ean
     Here
\ban
  Q^{L\prime\prime}_{a11}
        &=&  
	(\CN_{12} - {\tan\!\theta^{}_W} \CN_{11})
            \bigl[
                  U^S_{a1} \CN_{13} - U^S_{a2} \CN_{14}
            \bigr]
\\
       &+&  \sqrt2\lambda\CN_{15}
            \bigl[
                  U^S_{a1} \CN_{14} + U^S_{a2} \CN_{13}
            \bigr]
	    - 2\sqrt2 k U^S_{a3}\CN_{15}^2,
\\
\phi_{qL} &=& {\cal N}_{12} T_3 + {\cal N}_{11}(Q -T_3)\tan\theta_{W},
\\
\phi_{qR} &=& \tan\theta_{W}\  Q\  {\cal N}_{11},
\\
P_{q} &=&  \bigl(\frac{1}{2}+T_3\bigr) \frac{{\cal N}_{14}}{\sin\beta}
          + \bigl(\frac{1}{2}-T_3\bigr) \frac{{\cal N}_{13}}{\cos\beta}.
\ean

        The coefficients ${\cal A}_q$ and ${\cal C}_q$
	take into account squark mixing
   	$\tilde{q}_L-\tilde{q}_R$ and the contributions of
        all CP-even Higgs bosons.
	Under the assumption $\lambda=k=0$ these formulas coincide 
	with the relevant formulas in the MSSM 
\cite{DreesNojiriER}.
	In what follows we use notations and definitions of our papers 
\cite{Superlight}.	
 
\section{Constraints on the NMSSM parameter space}
\label{sec:constraints}
  	 Assuming that the neutralinos form a dominant part of
    	 the DM in the universe one obtains a cosmological constraint
   	 on the neutralino relic density.
   	 The present lifetime of the universe is at least $10^{10}$ years,
   	 which implies an upper limit on the expansion rate and
   	 correspondingly on the total relic abundance.
   	 The contribution of each relic particle species $\chi$  has to obey
~\cite{kolb}:
$
 \Omega_\chi h^2_0<1,
$
   	where the relic density parameter  $\Omega_\chi = \rho_\chi/\rho_c$
   	is the ratio of the relic neutralino mass  density $\rho_\chi$
   	to  the critical one
   	$\rho_c = 1.88\cdot 10^{-29}h^2_0~$g$\cdot$cm$^{-3}$.
	
	To have the papameter $\Omega_{\chi} h^2_0$ under 
	control
   	we calculate it
	following the standard
   	procedure on the basis of the approximate formula
\cite{DreesNojiriOM}:
$$
\Omega_{\chi} h^2_0 = 2.13\times 10^{-11}
\left(\frac{T_{\chi}}{T_{\gamma}}\right)^3
\left(\frac{T_{\gamma}}{2.7K^o}\right)^3
\times N_F^{1/2}
\left(\frac{{\mbox{GeV}}^{-2}}{a x_F + b x_F^2/2}\right).
$$
   	Here $T_{\gamma}$ is the present day photon temperature,
   	$T_{\chi}/T_{\gamma}$ is the reheating factor,
   	$x_F = T_F/m_{\chi} \approx 1/20$, $T_F$ is
   	the neutralino freeze-out temperature, and $N_F$ is the total
   	number of degrees of freedom at $T_F$. 
	The coefficients $a, b$ are determined from the
   	non-relativistic expansion
$
\langle \sigma_{ann.} v \rangle  \approx a + b x
$
   	of the thermally averaged cross section of 
	neutralino annihilation in the NMSSM.
   	We adopt an approximate treatment not taking into account
   	complications, which occur when the expansion fails
\cite{omega}. 
	We take into account all possible channels of
   	the $\chi$-$\chi$ annihilation in the NMSSM 
\cite{AStephan}.

   	Since the neutralinos are mixtures of gauginos,
   	higgsinos, and singlino 
	the annihilation can occur both, via
   	s-channel exchange of the $Z^0$ and Higgs bosons and
   	t-channel exchange of a scalar particle, like a 
	selectron. 
   	This constrains the parameter space, as discussed by
   	many groups
\cite{DreesNojiriOM,relic}. 

        In the analysis we ignore possible rescaling of the local
        neutralino density $\rho$ which may occur in the region of the
        NMSSM parameter space where $\Omega_\chi h^2_0 < 0.025$
\cite{GelminiGonfoloRoulet}--\cite{BBEFMS}.  
        If the neutralino is accepted as a dominant part of the DM its
        density has to exceed the quoted limiting value 0.025.
        Otherwise the  presence of additional DM components should be
        taken into account
	and the SUSY solution  of the DM problem
	with such low neutralino density becomes questionable.
        We assume neutralinos to be a dominant component of
        the DM halo of our galaxy with a density
        $\rho_{\chi}$ = 0.3 GeV$\cdot$cm$^{-3}$ in the solar vicinity
        and disregard in the analysis points with
        $\Omega_{\chi}h^2_0 < 0.025$.

	The parameter space of the NMSSM 
	is constrained by the non-observation of the 
	supersymmetry particles with  
	high energy colliders LEP and Tevatron.
	Due to the possible presence of singlino component,
	the $Z\tilde{\chi}^0\tilde{\chi}^0$ 
	and $Z S_a P_\alpha $ 
	couplings can be  suppressed in the NMSSM. 
	Therefore the neutralino and Higgs mass bounds are much weaker
	than in the MSSM
\cite{FrankeFraas}.
	The consequences from the negative neutralino search at LEP
	for the parameter space and the neutralino masses 
	have been studied in 
\cite{FrankeFraasBartl, FrankeFraas}.

	We used the following constraints from LEP.
	For new physics contributing to the total $Z$ width
$\Delta \Gamma(Z\to \tilde{\chi}^+\tilde{\chi}^- 
	+ Z\to \tilde{\chi}^0_i\tilde{\chi}^0_j) < 23$~MeV.
	For new physics contributing to the invisible $Z$ width
$\Delta \Gamma(Z\to \tilde{\chi}^0_i\tilde{\chi}^0_j) < 8$~MeV.
	From the direct neutralino search
$B (Z \rightarrow \tilde{\chi}^0_1 \tilde{\chi}^0_j) < 2 \times 10 ^{-5}$
	for $j=2,\ldots,5$, 
	and 
$B (Z \rightarrow \tilde{\chi}^0_i \tilde{\chi}^0_j)
	< 5 \times 10 ^{-5}$, for $i,j=2,\ldots,5$.
	We have also included 
	the experimental bounds from the direct search for
	pseudoscalar Higgs bosons produced 
	together with a Higgs scalar at LEP
\cite{ALEPH}, 
	but this does not significantly affect the excluded parameter domain
\cite{FrankeFraas}.

	In our numerical analysis we use the following 
	experimental restrictions for the SUSY particle spectrum in the
	NMSSM:
$ m_{\tilde{\chi}^+_1} \geq 90$~GeV for charginos,
$ m_{\tilde{\nu}}      \geq 80$~GeV for sneutrinos,
$ m_{\tilde{e}_R}      \geq 80$~GeV for selectrons,
$ m_{\tilde{q}}       \geq 150$~GeV  for squarks,
$ m_{\tilde{t}_1}      \geq 60$~GeV  for light stop,   
$ m_{H^+}              \geq 65$~GeV  for charged Higgses and 
$ m_{S_1}              \geq 1$~GeV 
	for the light scalar neutral Higgs.
	In fact, it appeared that all above-mentioned 
	constraints do not allow 
	$m_{S_1}$ to be smaller then 20 GeV.

\section{Numerical Analysis} \label{sec:numerical}

\begin{figure}[p]
\begin{picture}(100,160)
\put(-3,-5){\includegraphics{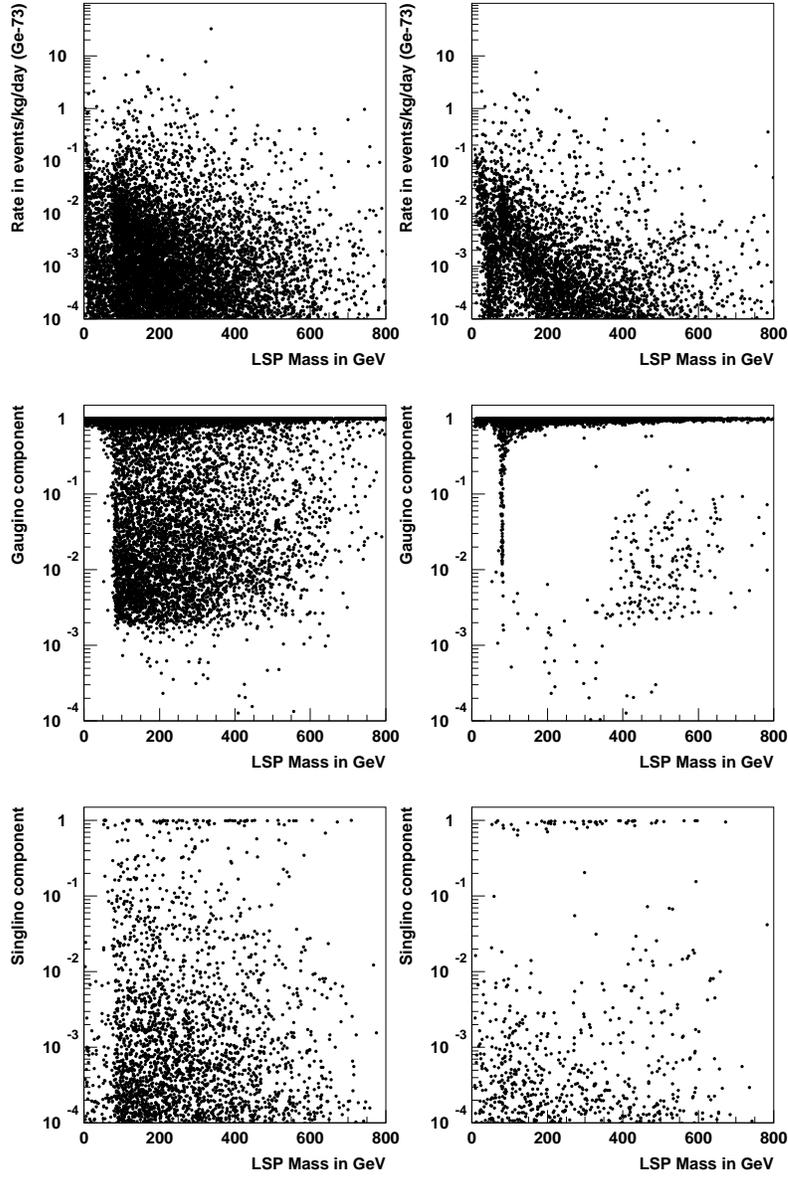}}
\end{picture}
\caption{
	Total event rate $R$ for
	$^{73}$Ge, the LSP gaugino 
	fraction ($\CN^2_{11}+\CN^2_{12}$) and
	singlino fraction ($\CN^2_{15}$)
	versus the LSP mass (from up to down).
	The left (right) panel presents results 
	obtained without (with) taking into account
	the cosmological relic density constraint.
}
\end{figure}

	Randomly scanned parameters of the NMSSM 
	at the Fermi scale are the following:
	the gaugino mass parameters $M^\prime$ and $M$, 
	the ratio of the doublet vacuum expectation values, $\tan\beta$, 
	the singlet vacuum expectation value $x$,
	the couplings in the superpotential $\lambda$ and $k$, 
	squared squark mass parameters $m^2_{Q_{1,2}}$ for the 
	first two generations and $m^2_{Q_3}$ for the third one,
	the parameters $A_\lambda$, $A_k$, 
	as well as $A_t$ for the third generation.
	The parameters are varied in the intervals given below
\begin{center}
\footnotesize\rm
\begin{tabular}{c@{~~$<$~~}c@{~~$<$~~}c}
\topline
  $-1000$~GeV  & $M^\prime$ &    1000~GeV   \\
  $-2000$~GeV  & $M$        &    2000~GeV   \\
        1      & ${\tan\!\beta}$&      50   \\
        0~GeV  & $x$        &   10000~GeV   \\
   $-0.87$     & $\lambda$  &    0.87   \\
   $-0.63$     & $k$        &    0.63   \\
  100~GeV$^2$  & $m^2_{Q_{1,2}} $    & 1000000~GeV$^2$  \\
  100~GeV$^2$  & $m^2_{Q_3} $    & 1000000~GeV$^2$  \\
  $-2000$~GeV  & $A_t$       &    2000~GeV   \\
  $-2000$~GeV  & $A_\lambda$ &    2000~GeV   \\
  $-2000$~GeV  & $A_k$       &    2000~GeV.   \\
\bottomline
\end{tabular}
\end{center}
	For simplicity 
	the other sfermion mass parameters 
	$m^2_{U_{1,2}}$, $m^2_{D_{1,2}}$, 
	$m^2_{L_{1,2}}$, $m^2_{E_{1,2}}$, and 
	$m^2_{T}$, $m^2_{B}$, 
	$m^2_{L_{3}}$, $m^2_{E_{3}}$, are chosen to be 
	equal to $m^2_{Q_2}$ and $m^2_{Q_3}$, respectively.
	Therefore masses of the sfermions in the same generation 
	differ only due to the D-term contribution.
	Other parameters (except $A_t$) of the supersymmetry 
	breaking potential $A_U$, $A_D$, $A_E$ 
	(for all three generations) are fixed to be zero.

   	The main results of our scan are presented in
	Figs. 1--4 in the form of scatter plots.
	Given in 
Fig.~1 
	are the total event rate $R$ for
	$^{73}$Ge, and the LSP gaugino 
	fraction ($\CN^2_{11}+\CN^2_{12}$) and singlino 
	fraction ($\CN^2_{15}$)
	versus the LSP mass.
	The left panel in 
Fig.~1 presents the above-mentioned
	observables obtained without taking into account
	the cosmological relic density constraint.
	In this case the total expected event rate $R$ reaches
	values up to about 50 events per day and per 
	1 kg of the $^{73}$Ge isotope. 
	As on can see from 
Fig.~1 
	the small-mass LSP (less then about 100~GeV) are mostly
	gauginos, with very small admixture of the singlino component.
	Large masses of the LSP (larger then 100~GeV)
	correspond to sizable gaugino and singlino components
	together with some higgsino fraction.
	The results of implementation of the
	cosmological constraint 
$$
\label{cosmology}
0.025 < \Omega_\chi h^2_0<1.
$$
	one can see in the right panel of 
Fig.~1.
	There is approximately a 5-fold reduction of the
	number of the points which fulfill all restrictions in this case.
	Nevertheless quite large values of event rate $R$
	(above 1 event/day/kg) still survive the cosmological
	constraint.
	The gaugino component becomes more significant, but
	the singlino fraction can not be completely ruled out
	especially for large masses of the LSP.
	The higgsino component of the LSP remains still possible
	only for LSP masses in the vicinity of $M_Z$.	
	The lower bound for the mass of the LSP now becomes
	about 5--7~GeV 
(Fig.~2).
\begin{figure}[h!]
\begin{picture}(100,45)
\put(-3,-65){\includegraphics{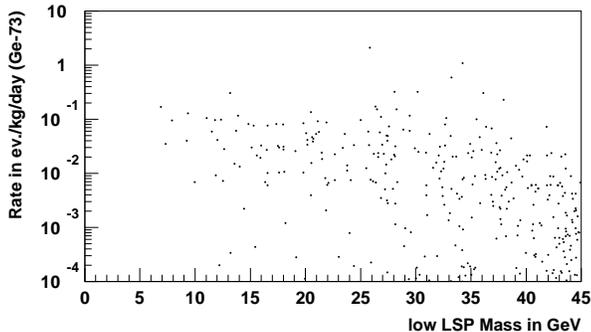}}
\end{picture}
\caption{
	Total event rate $R$ for
	$^{73}$Ge versus the LSP mass in the 
	region of low LSP mass.
}
\end{figure}

\begin{figure}[t!]
\begin{picture}(100,78)
\put(2,-50){\includegraphics{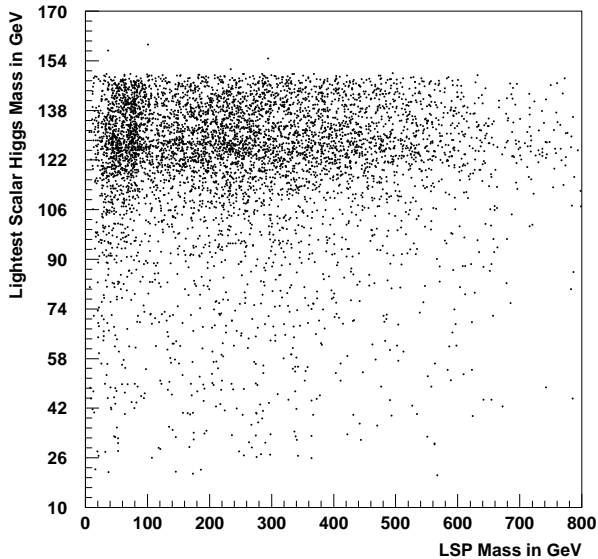}}
\end{picture}
\caption{Correlations between mass of the LSP and
	mass of the lightest scalar Higgs boson $S_1$.}
\end{figure}

Figure~3 shows correlations between mass of the LSP and
	mass of the lightest scalar Higgs boson $S_1$.
	The upper (about 160~GeV) and lower (about 20~GeV) 
	bound for the mass of $S_1$ are clear seen.	
	Given in 
Fig.~4 
	are the total event rate $R$ for
	$^{73}$Ge 
	versus the masses of all scalar Higgs bosons
	$S_{1,2,3}$. 
\begin{figure}[t!]
\begin{picture}(100,130)
\put(0,-20){\includegraphics{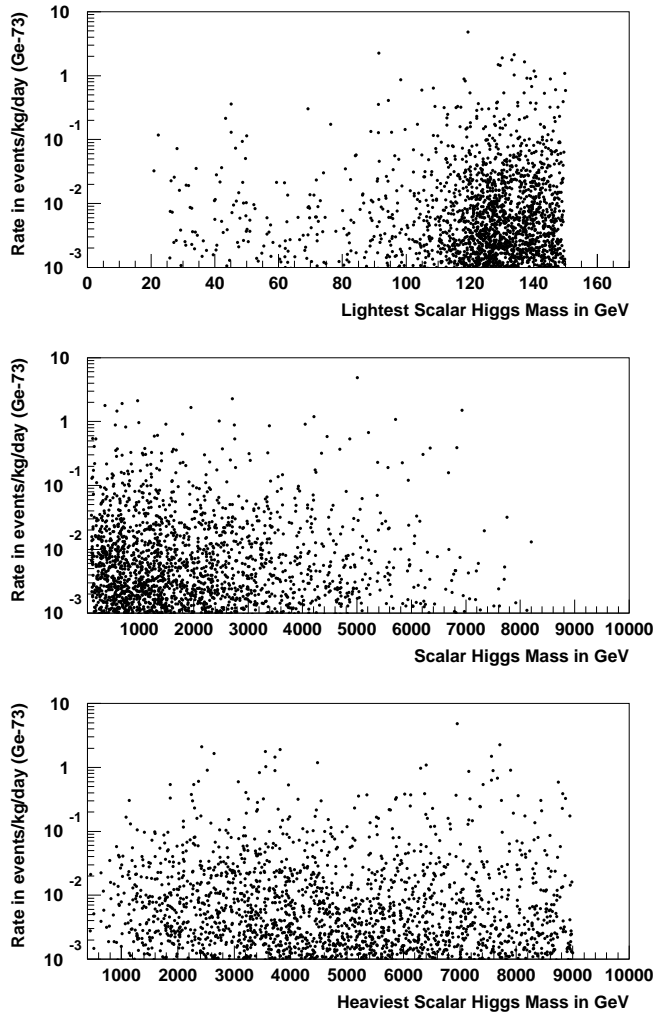}}
\end{picture}
\caption{
	Total event rate $R$ for
	$^{73}$Ge versus 
	versus the masses of scalar Higgs bosons
	$S_{1,2,3}$
	(from up to down).
}
\end{figure}

\section{Conclusion}  \label{sec:conclusion}
        In not too far future new very
        sensitive dark matter (DM) detectors
\cite{GENIUS}--\cite{Baudis}
	may start to operate, and 
        one expects new, very important data from these experiments.
        The prospects of the direct and indirect detections
        of the LSP have comprehensively been investigated
\cite{JungKamGriest} in the various versions of the MSSM.

	In the paper we address the question
	whether the Next-to-Minimal Supersymmetric Standard Model
	can be attractive from the point of view of the direct detection
	of the neutralinos provided the neutralino is 
	the stable LSP.
	To answer the question we derived
	the effective low energy neutralino-quark Lagrangian, 
	which takes into account the contributions of extra 
	scalar Higgs boson and extra neutralino.
	On this basis we calculated the total 
	direct-dark-matter-detection event rate in $^{73}$Ge as a  
	representative isotope which is interesting 
	for construction of a realistic dark matter detector.	
	We analyzed the NMSSM taking into account
	the available accelerator and cosmological constraints
	by means of random scan of the NMSSM parameter space 
	at the Fermi scale.
	We demonstrated that the cosmological constraint does not 
	rule out domains in the parameter space which correspond to 
	quite sizable event rate in a germanium detector.

	Due to relaxation of the gaugino unification condition
	we found domains in the parameter
	space where lightest neutralinos have quite small masses
	(about 7~GeV), acceptable relic
	abundance and sufficiently large expected event rate
	for direct detection with a $^{73}$Ge-detector.

	Therefore the NMSSM looks not worse then the MSSM from the
	point of view of direct dark matter detection.
	The question arises:
	Is it possible to distinguish MSSM and NMSSM 
	by means of direct dark matter detection of LSP? 
	It is a problem to be solved in future.
	The question can disappear by itself
	if negative search for light Higgs with LHC 
	rules out the MSSM.
	In fact, 
	the upper tree level mass bound for the
	lightest Higgs scalar of the MSSM
\begin{equation} \label{tlb}
	m_h^2 \leq m_Z^2 \cos^22\beta
\end{equation}
	is increased in the NMSSM to
$ m_{S_1}^2 \le m_Z^2 \cos^2 2\beta+\lambda^2(v_1^2+v_2^2) \sin^2 2\beta$
	and the NMSSM may still remain a
        viable model when the MSSM can be ruled out due to 
(\ref{tlb}).
	Therefore the NMSSM might remain a viable theoretical 
	background for direct dark matter search for relic
	neutralinos in the post-MSSM time.

\section*{Acknowledgments}
	We thank S.G. Kovalenko for helpful discussions.
	The investigation was supported in part (V.A.B.)
	by Grant GNTP 215 from Russian Ministry of Science
	and by joint Grant 96-02-00082 from Russian Foundation 
	for Basic Research and Deutsche Forschungsgemeinschaft.


\end{document}